\documentclass[amsmath,amssymb]{revtex4}
\usepackage{epsfig}
\begin{document}

\title[]{Braneworld with Induced Axial Symmetry}

\author{Edgard Casal de Rey Neto}

\email{{eneto@ita.br}}
\affiliation{Instituto Tecnol\'ogico de Aeron\'autica - Divis\~ao de Ensino Fundamental\\ Pra\c ca Marechal Eduardo Gomes 50,~S\~ao Jos\'e dos
Campos, 12228-900 SP, Brazil}

\begin{abstract}
We take arbitrary gravitational perturbations of a 5d spacetime and reduce it to the form an axially symmetric warped braneworld. Then, we write the filed equations for the linearized gravity perturbations. We obtain the equations that describes the graviton, gravivector and the graviscalar fluctuations and analyse the effects of the Schr\"odinger potentials that appear in these equations.

\end{abstract}

\maketitle


\section{Introduction}

In the almost all works on the Randall-Sundrum (RS) braneworlds~\cite{RS12}, the axial gauge is used to derive the linearized gravity dynamics. In this gauge there are no fluctuations transverse to the brane and, the scenario is axially symmetric. An other important gauge is the harmonic (de Donder) gauge for which, in 5D, the $h_{55}$-graviscalar and the $h_{5\mu}$-gravivector 4D fluctuations can be non zero, breaking the axial symmetry. However, as pointed out in~\cite{PL1}, a new coordinate frame in 5D can be found, where the metric becomes axially symmetric even with $h_{55}$, $h_{5\mu}\neq 0$. By following~\cite{PL1}, we call such coordinate frame the {\it local frame}. Here, we analyse the field equations for the vacuum fluctuations that arise in the local frame, using  the 5D de Donder gauge.

\section{Axially symmetric braneworld}
\noindent
The 5D metric is expanded as $g_{AB}=\eta_{AB}+h_{AB}$ where $A,B=0,1,2,3,5$, $\eta_{AB}=diag(-1,1,1,1,1)$ and $h_{AB}$ small gravity fluctuations. The 5D line element is
\begin{equation}
\label{5Dle}
dS^2=(\eta_{\mu\nu}+h_{\mu\nu})dx^\mu dx^\nu+2h_{5\mu}dx^\mu dx^5+(1+h_{55})(dx^5)^2.
\end{equation}
The above spacetime take the axially symmetric form
\begin{equation}
ds^2=g_{(\mu)(\nu)}dx^{(\mu)}dx^{(\nu)} +\epsilon^2 dy^2
\end{equation}
where $x^{(A)}=e^{(A)}_B x^B$,
with the $e^{(A)}_B$ given by
\begin{equation}
e^{(\mu)}_A=\delta^\mu_A,\quad e^{\mu}_{(\nu)}=\delta^\mu_\nu,\quad e^\mu_{(5)}=0,\quad e^{(5)}_\mu=\epsilon^{-1} P_\mu,
\end{equation}
\begin{equation}
e^{5}_{(\mu)}=-P_\mu,\quad e^{(5)}_5=\frac{1+\varphi/2}{\epsilon},\quad e^5_{(5)}=\epsilon(1-\varphi/2),
\end{equation}
where $P_\mu= h_{5\mu}$, $\varphi= h_{55}$, $y=x^{(5)}$ and  $\epsilon^2=-1,1$. The physical 4D metric can be given by $g_{(\mu)(\nu)}$, or  $\hat{g}_{(\mu)(\nu)}=e^{-2f}g_{(\mu)(\nu)}$. If $f=f(y)$, we have
\begin{equation}
\label{inducedbw}
ds^2=e^{-2f(y)}g_{(\mu)(\nu)}dx^{(\mu)}dx^{(\nu)}+\epsilon^2 dy^2.
\end{equation}
Then, we assume that~(\ref{inducedbw}) satisfies the action
\begin{equation}
\label{action}
S\!=\!\!\int\!{d^5x[\sqrt{-g}(\kappa^{-2}R\!+\!\Lambda_5\!+\!{\cal L}_m)+\!\sqrt{-g_b}\sigma]},
\end{equation}
where $g={\rm det}(g_{{(A)(B)}})$, $g_b=det(g_{(\mu)(\nu)})$ and $\kappa^2=M_*^{-3}$. The vacuum solution gives $f'(y)^2=-\kappa^2\Lambda_5(y)/12$ and $f''(y)=\kappa^2\sigma(y)/12$.
\section{Local frame gravitational fluctuations}
\noindent
We work in the conformal frame where
\begin{equation}
\label{warpsym}
ds^2=e^{-f(z)}[g_{(\mu)(\nu)}dx^{(\mu)}dx^{(\nu)}+\epsilon^2 dz^2].
\end{equation}
The equations for the gravitational fluctuations are derived with $g_{(\mu)(\nu)}=e_{(\mu)}^Ae_{(\nu)}^B g_{AB}$ and,
$\bar{\partial}_{(A)}(...)=e^B_{(A)}\partial_B (...)$, with $[\bar{\partial}_{(A)},\bar{\partial}_{(B)}]\neq 0$. The $h_{AB}$ satisfies $\partial^B h_{AB}=0,\quad h_C^C=0$,
which means that
\begin{equation}
\label{gaugeeq}
h_{\alpha}^\alpha=-\varphi,\quad \partial^\alpha{P_\alpha}=-\varphi',\quad \partial^\alpha h_{\mu\alpha}=-P'_{\mu}.
\end{equation}
The the ``prime'' represents $\partial_z$.
The field equations are
\begin{equation}
\label{scaleq}
\Box  \varphi+12 f'^2\varphi+3f'\varphi'=-\kappa T^{(m)}_{55},
\end{equation}
\begin{equation}
\label{veceq}
\Box P_\mu+6(f'^2+f'')P_\mu+3f'\partial_\mu \varphi-2[\bar{\partial}_{(\mu)},\bar{\partial}_{(5)}]f=-\kappa T^{(m)}_{\mu 5},
\end{equation}
\begin{eqnarray}
\label{tenseq}
&\Box&\!\!h_{\mu\nu}+3f'(\partial_\mu P_\nu+\partial_\nu P_\mu)-3f'h'_{\mu\nu}+12 (f'^2-f'')\varphi\eta_{\mu\nu}\nonumber \\
&-&2[\bar{\partial}_{(\mu)},\bar{\partial}_{(\nu)}]f=-\kappa T^{(m)}_{\mu\nu},
\end{eqnarray}
where $\Box=\eta^{AB}\partial_A\partial_B$. The local frame equations
depends on the comutator of partial derivatives of the warp function. The equation for the scalar do not changes~\cite{dDrefs}.\\
\noindent
{\bf Extended KK-gravity: $\Lambda_5=\sigma=0$}.
The system~(\ref{scaleq})-(\ref{tenseq}) decouples to
\begin{equation}
\label{dcsys1}
\Box \phi=-\kappa T^{(m)}_{55},\quad\Box P_\mu=-\kappa T^{(m)}_{5\mu},
\end{equation}
\begin{equation}
\label{dcsys3}
\Box h_{\mu\nu}=-\kappa T^{(m)}_{\mu\nu}.
\end{equation}
The scenario is an extended Kaluza-Klein gravity with a no compact extra dimension. The gauge conditions enable us to write the 4D tensor $h_{\mu\nu}$ in terms of spin-2, spin-1 and spin-0 fluctuations. The vacuum is flat.
\section{Warped geometry fluctuations}
\noindent
{\bf Graviscalar}. Consider $T^{(m)}_{55}=0$. The $\varphi$ is re-scaled to $\varphi(x,z)=e^{-3f(z)/2}\tilde{\varphi}(x,z)$ and we look for solutions $\tilde{\varphi}(x,z)=\tilde{\varphi}(x)\psi_s(z)$, with $\partial^\alpha\partial_\alpha\tilde{\varphi}(x)=m_{s}^2\tilde{\varphi}(x)$. Then,~(\ref{scaleq}) implies
\begin{equation}
\label{scheqscal}
[-\partial_z^2+V_{s}(z)]\psi_s(z)=m_s^2\psi_s(z),
\end{equation}
with $V_s(z)=-(\frac{3}{2}f''-\frac{39}{4}f'^2)$.
The vacuum solutions for $\varphi$ are of no interest, because the compatibility condition between~(\ref{scaleq}) and~(\ref{veceq}) force us to set $\varphi=0$~\cite{dDrefs}.\\
\noindent
{\bf Gravivector.} With $\varphi=0$, eq.~(\ref{veceq}) in vacuum gives
\begin{equation}
\label{scheqvec}
[-\partial_z^2+V_v(z)]\psi_v(z)=m_v^2\psi_v(z),
\end{equation}
where $V_v(z)=-5(f'^2+f'')$,
for $\psi_v$ defined by $\tilde{P}_\mu(x,z)=\tilde{P}_\mu(x)\psi_v(z)$,
$\partial^\alpha\partial_\alpha\tilde{P_{\mu}}(x)=m_{v}^2\tilde{P_\mu}(x)$,and
$\tilde{P}_\mu(x,z)=e^{f(z)}{P_\mu}(x,z)$.
For RS warp, the potential is $V_v(z)=-10k\delta(z)$. Then, we have masive solutions with $m_v^2=-25 k^2$, whereas $m_v^2=-36k^2$ in the coordinate frame~\cite{dDrefs}.

The compatibility condition between~(\ref{veceq}) and~(\ref{tenseq}) is
\begin{equation}
\label{compcond1}
[-26(f'^3+f'f'')+6 f''']\tilde{P}_\mu+(8f''-4f'^2)\tilde{P}'_\mu=0.
\end{equation}
If $f={\rm Log}(k|z|+1)$, we have
\begin{eqnarray}
\label{compcond2}
\Big(&-&10k\delta(z){\rm sgn}(z)+3k\delta'(z)+3\frac{k^3{\rm sgn}(z)}{(k|z|+1)^3}\Big)\psi_v(z)\nonumber \\
&+&\Big(4k\delta(z)-3\frac{k^2}{(k|z|+1)^2}\Big)\psi'_v(z)=0.
\end{eqnarray}
For $|z|\geq 0$,~(\ref{compcond2}) is satisfied by $\psi_v=25a_v(k|z|+1)$,
where $a_v$ is a constant. At $z=0$,~(\ref{compcond2}) implies, $a_v=0$.
With the smooth warp, $f(z)={\rm Log}(k^2z^2+1)$, eq.~(\ref{compcond1}) becomes
\begin{equation}
\label{compcond3}
\left[\frac{6\,k^5\,z^3}{{\left( 1 + k^2\,z^2 \right) }^2}\!-\!
   \frac{11\,k^3\,z}{{1 + k^2\,z^2}}\right]\psi_v-
\left[\frac{3k^3z^2}{1+k^2z^2}-1\right]\psi'_v=0.
\end{equation}
At $z=0$, we have $\psi'_v(0)=0$.
A solution of~(\ref{scheqvec}) that satisfy~(\ref{compcond3}), is the massive mode described by
\begin{equation}
\label{psivsol}
\psi_v=c_v\frac{e^{\frac{13}{4}{\;\tanh}^{-1}(1/3+4k^2z^2/3)}}{(-1+k^2z^2+2k^4z^4)^{5/8}},
\end{equation}
with mass $m_v^2=-{3k^2\left( 7 + 25k^2z^2 \right) }/
 {\left( 1 - 2k^2z^2 \right)}^2$.

The figure 1 shows the variation of $m_v^2 $ with the extra coordinate. It is almost constant for small $z$ and diverges for $z\rightarrow \pm z^*=\pm 1/(\sqrt{2}k)$. On the $|z|=0$ 3-brane,
$\tilde{P}_\mu(x)=c_\mu e^{\frac{13}{4}{\;\tanh}^{-1}(1/3)}e^{ip_\alpha x^\alpha}e^{-i\frac{5\pi}{8}}$,
where $p^2=m_v^2(0)=-21k^2$ and, $p^\mu c_\mu=0$.

\setcounter{figure}{0}
\begin{figure}
\begin{center}
\epsfig{figure=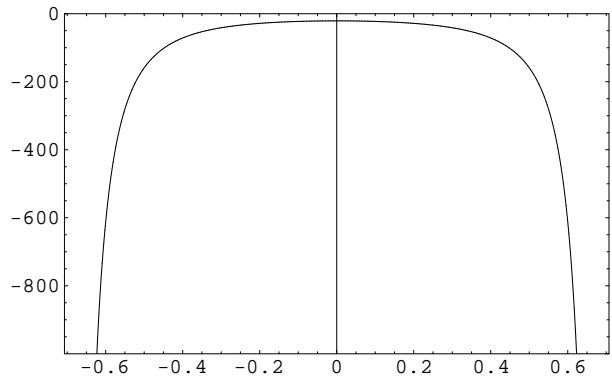,angle=360,height=5cm,width=8.0cm}
\caption{\label{fig1}The squared tachyonic mass, $m_v^2(z)$, in the vertical  axis. The horizontal axis is $z^*\leq z\leq z^*$ and $k=1$.}
\end{center}
\end{figure}
\noindent
{\bf Graviton}. To obtain the graviton potential we take $c_\mu=0$ and $T^{(m)}_{\mu\nu}=0$. Then, the eq.~(\ref{tenseq}) implies
\begin{equation}
\label{scheqtens}
[-\partial_z^2+V_g(z)]\psi_g(z)=m_g^2\psi_g(z),
\end{equation}
where $V_g=-(\frac{3}{2}f'^2-\frac{9}{4}f'')$. This potential reproduces the RSII result for $f(z)={\rm Log}(k|z|+1)$.
\section{Conclusions}
In the no warped scenario, the vacuum fluctuations are described by three independent wave equations which describes the 4D scalar, vector and tensor fluctuations on the $|z|\!\!=\!\!0$ 3-brane. In warped scenarios, there are no scalar propagation on the 3-brane vacuum. For the RS warp, there are no gravivector on the 3-brane. For the smoothed warped braneworld, we obtain a tachyonic mass solution for the gravivector, that also satisfies the compatibility condition. This solution becomes a massless spin-1 fluctuation if ${\Lambda_5}\rightarrow 0$.
\section{Acknowledgements}
I  thank Dr. R. M. Marinho Jr. This work supported by brazilian agency CNPQ (gr. 150854/2003-0) and ITA.

\end{document}